\documentclass[prc,twocolumn,showpacs,floatfix,nofootinbib,preprintnumbers,superscriptaddress,amsmath,amssymb]{revtex4}

\usepackage{CJKutf8}
\usepackage{graphicx}
\usepackage{epsfig}
\usepackage{amsmath,amssymb}
\usepackage{bm}
\usepackage{color}
\usepackage{float}
\usepackage{dcolumn}
\usepackage{multirow}
\usepackage{mathrsfs}
\usepackage[arrowdel]{physics}
\usepackage[OT1]{fontenc}
\usepackage{outline}
\usepackage{tikz,tikz-3dplot}
\usetikzlibrary{positioning}

\DeclareMathAlphabet{\mathcal}{OMS}{cmsy}{m}{n}

\begin{document}
\begin{CJK*}{UTF8}{gbsn}

\title{A cranked self-consistent mean-field description of the triaxially deformed rotational bands in $^{138}$Nd}

\author{Yue Shi (石跃)}
\email{yueshi@hit.edu.cn}
\affiliation{Department of Physics, Harbin Institute of Technology, Harbin 150001, People's Republic of China}

\begin{abstract}
\begin{description}
\item[Background] Compared to the axially deformed nuclei, triaxially deformed
ones are relatively scarce. This is mainly due to the difficulties in the
identification of experimental signatures pertaining to the triaxial degree of
freedom. In the nucleus $^{138}$Nd, a number of rotational bands have been
observed to have medium or high spin values. They have been 
interpretated in the macroscopic-microscopic method, to be based
on triaxial minima. In particular, for a few configurations, the calculations
suggested that a re-orientation of the rotational axis may have occurred along
the rotational bands. 

\item[Purpose] The present work aims at a quantitative description of the
experimentally observed bands in $^{138}$Nd, using the cranked self-consistent
Skyrme-Hartree-Fock (SHF) method or cranked nuclear density functional theory (DFT). 
Such a study, which is still missing, will
provide alternative interpretations of the structure of the bands and hence, shed
new lights on the triaxiality issue in connection with experimental data.

\item[Methods] The rotational bands are described using cranked self-consistent
mean-field method with SLy4 and SkM* Skyrme energy density functionals (EDFs). 
For SLy4 EDF, the time-odd pieces are included using Landau parameters (denoted 
with SLy4$_{\rm L}$). For SkM* EDF, the local gauge invariance argument has been used to
determine the time-odd components of the mean-field.

\item[Results] The survey of different configurations near Fermi surface of
$^{138}$Nd results in 12 lowest configurations, at both positive- and
negative-$\gamma$ deformations. These are calculated to be the energetically 
lowest configurations. The results show that, for both EDFs, the rotational 
states based on positive-$\gamma$ minimum, which is at $\gamma\approx35^{\circ}$, 
are lower than the respective configurations with negative-$\gamma$ deformation. 
The general trends of the spin-versus-omega curve, and
the energy-versus-spin curve reproduce well those of the experimental data. 
Further, for the observed bands `T1-T8', the calculated results using 
SLy4$_{\rm L}$ allows the configurations of the observed bands to be 
assigned. The calculations predict transitional quadrupole moments, 
which can be used to compare with future experimental data.

\item[Conclusions] The current cranked self-consistent mean-field calculations of the
near-yrast high-spin rotational bands in $^{138}$Nd reproduce well
the experimental data. The results suggest that the experimentally observed
bands can be assigned to the calculated bands with various configurations at
the positive-$\gamma$ deformation. The predictions of the current calculations are 
complementary to that of the well-know macroscopic-microscopic calculations, 
both of which await future experiment to verify.

\end{description}
\end{abstract}


\maketitle
\end{CJK*}

\section{Introduction}
\label{intro}

It has been known that the occurrence of rotational bands in a nucleus is due
to the break of rotational invariance of the nuclear mean-field Hamiltonian in
the intrinsic frame~\cite{bohr75,nils95}. There are a few fascinating phenomena
which can be associated with the triaxial deformation of a nucleus. Among them,
the energy inversion of the signautre partners, differing from a normal order
predicted by rotational models without the triaxial deformation~\cite{beng84};
the occurrence of identical bands which are presumably due to the break of
chiral symmetry in a triaxially deformed mean-field~\cite{frau97,frau01}; as
well as the bands based on wobbling motions of a rotating nucleus~\cite{odeg01}
are perhaps the most significant examples.

Compared to the axially symmetric deformation, where the characterization has
been fairly well established both experimentally and theoretically, physics
associated with triaxially deformed nuclei has been rather scarce. This is due
to the more subtle experimental signatures associated with the triaxial degree
of freedom. Indeed, alternative explanations may co-exist to explain those
phenomena postulated to be due to triaxial deformation. In particular, for
states near ground states, a triaxially deformed nucleus tend to have soft
minima in the potential-energy-surface, where ambiguities arises as to weather
the observed triaxial deformations are of a static character.

Recently, in the Er-158 nuclear region, a number of strongly triaxially
deformed bands (TSD) have been observed to have large dynamic moments of
inertia, marking the return of collectivity in these nuclei, in which
core-breaking excitations were
observed~\cite{paul07,olli09,must11,olli11,agui08}. In the
macroscopic-microscopic calculations~\cite{beng85}, these bands were calculated
to be associated with triaxially deformed minima. In the two triaxially
deformed minima, the tilted-axis-cranked self-consistent mean-field
calculations~\cite{shi12,shi13} suggested that the energetically higher minimum
is unstable with respect to a re-orientation of the rotational axis. However,
these strongly triaxial deformed bands were not linked to the lower-energy
states where spin and energy values are known, leaving the spin and energy
values of these TSD bands underdetermined. This makes the interpretations of
the theoretical models uncertain~\cite{afan12}.

A much better laboratory for studying the triaxially deformed nucleus may be
$^{138}$Nd. Indeed, in this nucleus, a multitude of rotational bands have been
observed~\cite{petr99,petr12,petr13,petr15}. For them, the
macroscopic-microscopic model~\cite{beng85} has predicted a rather pronounced
static triaxial deformation. In particular, the linking transitions from these
band members to the energetically known lower-spin states have been observed
through precise $\gamma$-ray measurements. Consequently, the spin values and
the excitation energies of the bands have been determined, which provide much
better testing ground for theoretical models compared to the situation of
$^{158}$Er.

Incidentally, the macroscopic-microscopic calculations in Ref.~\cite{petr13}
showed that the structural changes in a pair of the rotational bands in
$^{138}$Nd were due to the deformation change from a positive-$\gamma$ value to
a negative-$\gamma$ one. These findings may constitute the first observation of
the re-orientation effect of the rotational axis with increasing rotational
frequency. It is then desirable to see the predictions from alternative
theoretical models, such as the self-consistent mean-field
methods~\cite{bender03,afan12}, which is still missing.

The present work aims at a description of the observed bands in
Refs.~\cite{petr12,petr15} using a cranked nuclear density functional theory
(DFT). The alternative description of the observables is supposed to shed new
lights on the underlying structures of these bands, leading to a better
understanding of triaxial deformation in nuclear structure. In
section~\ref{model}, I describe the model, and the parameter used in the
current work. Section~\ref{results} presents detailed results and discussions,
before a summary which is presented in section~\ref{summary}.

\section{The model}
\label{model}

In the present work, the cranked SHF calculations are performed with
symmetry-unrestricted solver HFODD (version 2.49t~\cite{doba09}). In this
model, the total energies (or Routhians in the rotational frame~\cite{olbr04})
are represented as functional of various densities~\cite{bender03}. The
single-particle Hamiltonian is expanded in terms of 969 deformed harmonics
oscillator basis, which are $\hbar\omega_x=\hbar\omega_y=0.4945$, and
$\hbar\omega_z=0.4499$\,MeV.

The SkM*~\cite{bart82} and SLy4~\cite{chab98} energy density functionals
(EDFs) are used in the particle-hole channel of the cranked SHF problem. For
cranking calculations, one deals with non-zero time-odd densities and fields.
The coupling constants are determined using local gauge invariance
arguments~\cite{doba95}, and Landau parameters~\cite{bender02}, for SkM* and
SLy4 EDFs, respectively. The latter is then denoted with SLy4$_{\rm L}$. The
current choice of the parameterizations including the time-odd part is 
identical with those appeared in Ref.~\cite{shi12,shi13} for consistency.

To obtain the solutions corresponding to the local minima that are interesting
for the current work, calculations with constraints on the quadrupole moments,
$Q_{20}$ and $Q_{22}$, are first performed. The constraints on $Q_{20}$ and
$Q_{22}$ deformations are then removed by zeroing the respective Lagrangian
multipliers. The earlier constraint solutions are used to warm start the
deformation unconstrained calculations. 

In the current principal-axis-cranking calculations, parity and signature
symmetries are always enforced. Each single-particle level could be labeled by
parity and signature quantum numbers. For details, see the caption of
Table~\ref{table1}.

Before I end this section, some explanations about (1) the sign of the $\gamma$
value which is used to measure the degree of triaxial deformation, and (2) the
relation between the commonly accepted convention (Lund convention) and the one
used in the current work is necessary. 

In the Lund convention, the rotation is chosen to be around the x-axis and the
deformation with $\gamma$ value in the interval of (0,60$^\circ$) corresponds
to a triaxially deformed nucleus that rotates around its short
axis~\cite{nils95}.

In the HFODD code~\cite{olbr04}, the one-dimentional cranking is around the
y-axis, and the $Q_{22}$ value is defined as
$\sqrt{3}\langle\hat{x}^2-\hat{y}^2\rangle$. In the current calculation, the
minima with positive $Q_{22}$ values are identified as minima with positive
$\gamma$ values. This definition is consistent with the Lund convention in the
fact that, for both conventions, a positive $\gamma$ value indicates rotation
around its short axis. 

To fully conform with the Lund convention which uses x-axis as the cranking
axis, a final step one needs to do is the following relabelling of the axes in
the HFODD code: (x,y,z)$\rightarrow$(y,x,-z). This results in the change of the
sign of $Q_{22}$ value. Such a convention transformation in our calculations
has been carefully considered. Note the negative sign in front of $Q_{22}$ in
Figure 3 of Ref.~\cite{shi12}, and in the current work.


\section{Results and discussions}
\label{results}

\begin{table}[htb]
\caption{The SHF configurations in $^{138}$Nd studied in this work. Each configuration
is described by the number of states occupied in the four parity-signature ($\pi$,$\rho$)
blocks, in the convention defined in Ref.~\cite{doba97}. 
For configurations `07' and `08', the `18$\rightarrow$19' means that the 
neutron is moved from the 18th to the 19th level. The transition quadrupole 
moment, $Q_t$ values, are calculated using the SLy4$_{\rm L}$ EDF 
at $\hbar\omega=0.6$\,MeV for the positive-$\gamma$ deformed minimum. 
The $Q_t$ values are calculated through the relation 
$Q_t=Q_{20}^p+\sqrt{\frac{1}{3}}Q_{22}^p$.}
\label{table1}
\begin{ruledtabular}
\begin{tabular}{ccccc}
Label & Configuration & $\pi$ & $\rho$ & $Q_t$ ({\it e}b) \\
\hline
01  &   $\nu$[20,20,19,19]$\otimes$$\pi$[16,16,14,14]   & +  & +1  & 2.7  \\
02  &   $\nu$[19,21,19,19]$\otimes$$\pi$[16,16,14,14]   & +  & $-$1  & 3.2  \\
03  &   $\nu$[20,20,19,19]$\otimes$$\pi$[16,15,15,14]   & $-$  & $-$1  & 3.0  \\
04  &   $\nu$[20,21,19,18]$\otimes$$\pi$[16,15,15,14]   & +  & $-$1  & 3.5  \\
05  &   $\nu$[21,20,18,19]$\otimes$$\pi$[16,16,14,14]   & $-$  & +1  & 2.6  \\
06  &   $\nu$[21,20,19,18]$\otimes$$\pi$[16,16,14,14]   & $-$  & $-$1  & 2.6  \\
07  &   $\nu$[21,20,18$\rightarrow$19,19]$\otimes$$\pi$[16,16,14,14]   & $-$  & +1  & 2.2  \\
08  &   $\nu$[21,20,19,18$\rightarrow$19]$\otimes$$\pi$[16,16,14,14]   & $-$  & $-$1  & 2.3  \\
09  &   $\nu$[20,20,19,19]$\otimes$$\pi$[15,16,14,15]   & $-$  & $-$1  & 2.7  \\
10  &   $\nu$[20,21,19,18]$\otimes$$\pi$[16,16,14,14]   & $-$  & +1  & 3.0  \\
11  &   $\nu$[20,21,18,19]$\otimes$$\pi$[16,16,14,14]   & $-$  & $-$1  & 3.0  \\
12  &   $\nu$[20,20,19,19]$\otimes$$\pi$[16,15,14,15]   & $-$  & +1  & 3.0  \\
\end{tabular}
\end{ruledtabular}
\end{table}

Table~\ref{table1} lists the configurations in $^{138}$Nd studied in this work.
Note that for each configuration, there are two states corresponding to the
minima with positive and negative $\gamma$ values. In the following
discussions, the `01' configuration with a positive-$\gamma$ minimum is denoted
with `pg01', the negative-$\gamma$ one is denoted with `ng01'. The same rule of
naming is applied to other configurations.

\begin{figure}
\centering
\includegraphics[scale=1.2]{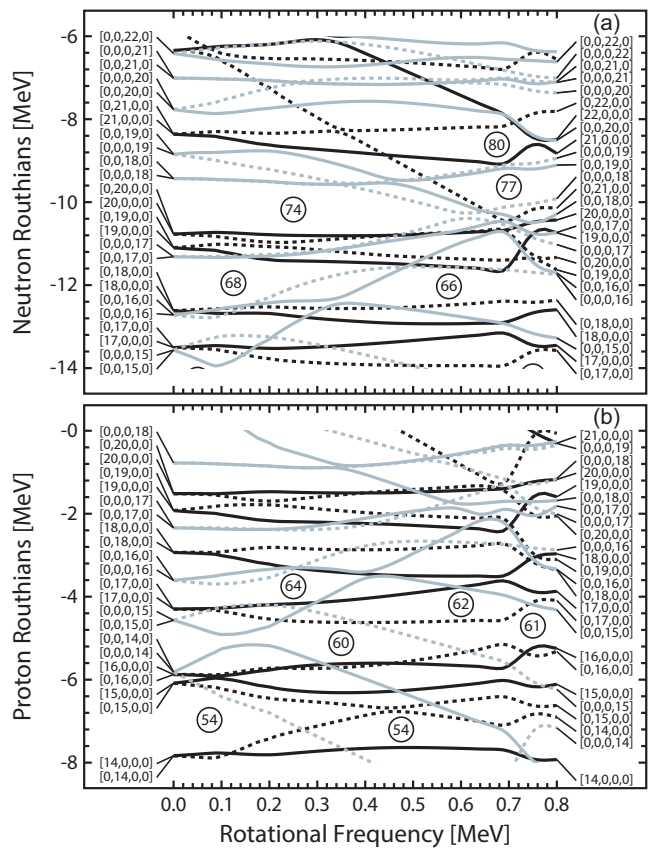}
\caption{Single-particle routhians from cranked SHF calculations 
with SLy4$_{\rm L}$ EDF for $^{138}_{60}$Nd$_{78}$. The configuration is `pg01' in 
table~\ref{table1}. The quadrupole moments for rotational frequencies
between 0.4-0.8\,MeV are ($Q_{20},Q_{22}$)$\approx$(10.0,5.3)\,b.
The levels with positive and negative parities are indicated by
black and light gray lines, respectively. The levels with $+$i and $-$i signatures
are indicated by solid and dashed lines, respectively.
One could find the order of the level within its $(\pi,\rho)$ block
at the beginning and the end of each line.}
\label{figure1}
\end{figure}

In figure~\ref{figure1}, I show the calculated single-particle routhians for
SLy4$_{\rm L}$ calculated with configuration `pg01' listed in
table~\ref{table1}. It can be seen that the configuration `pg01' contains no
particle-hole excitations for the frequency interval of $\sim$0.0-0.5\,MeV. It
has to be noted that, even though the configuration `pg01' appears to be the
lowest energetically, since the single-particle levels below the Fermi surfaces
of protons and neutrons are occupied, it does not necessarily indicate that the
obtained total energy of this configuration in the rotating frame is the
lowest. This is so because a state with another configuration may acquire
additional binding through self-consistent process. In addition, a state with
another configuration may be lower in energy by having a slightly different
deformation compared to the configuration `pg01'.

Nevertheless, it is reasonable to state that configuration `pg01' is among the
lowest in total Routhian for frequency interval of $\sim$0.0-0.5\,MeV, which
will be shown in the energy-versus-spin plot in later discussions. In
figure~\ref{figure1}, it can be seen that, for $\hbar\omega>0.5$\,MeV, a state
with $(\pi,\rho)=(+,-i)$ penetrates below the Fermi surface, and crosses with
the 20th level in the (+,$-$i) block. This crossing corresponds to the
characteristic backbending observed in the experimental data which will be
discussed later. The behaviors of spin and energy values as a function of
rotational frequency of configuration `pg01' are rather typical. Hence, the
quality of the description of these behaviors are particularly important for
assessing the usefulness of the current calculations.

\begin{figure}
\centering
\includegraphics[scale=0.3]{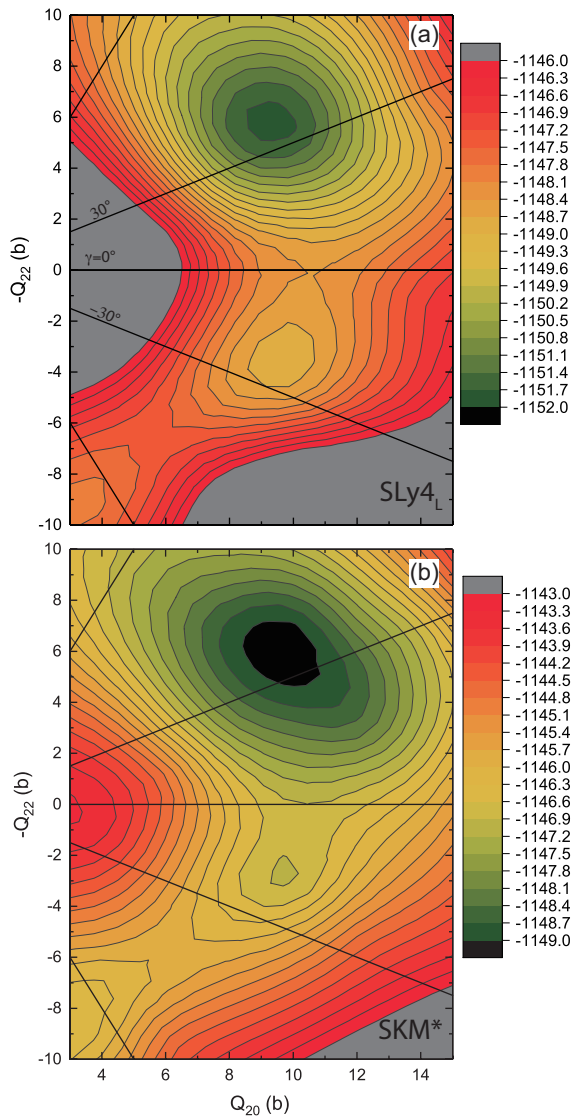}
\caption{Total Routhian surfaces for $^{138}$Nd for configuration 
`pg01' in table~\ref{table1}. Calculations are performed with Skyrme 
EDF SLy4$_{\rm L}$ (a), and SkM* (b) at a rotational 
frequency of 0.6\,MeV. Contour lines are 0.4\,MeV apart in energy.}
\label{figure2}
\end{figure}

Before showing the calculated total angular momenta and energies as a function
of rotational frequency, in figure~\ref{figure2}, I show the
total-Routhian-surfaces of $^{138}$Nd for configuration `01' in
table~\ref{table1}, calculated with SLy4$_{\rm L}$~\cite{chab98} (a),
and SkM*~\cite{bart82} (b) EDFs, at a rotational frequency of
0.6\,MeV. For both EDFs, the minima at the positive and negative $\gamma$
deformations can be seen. The minima with positive $\gamma$ values are deeper,
and are larger in terms of $|\gamma|$ value. For SLy4$_{\rm L}$, the
positive-$\gamma$ minimum is deeper than that of the negative-$\gamma$ one by
$\sim$3.0\,MeV in energy. Whereas the energy difference for the two minima of
SkM* is only $\sim$1.5\,MeV. The barrier separating the two minima in the
$\gamma$ direction is lower for SkM* compared to that of SLy4$_{\rm L}$. Note
that for SkM* EDF, the crossing with the lowest level from $\mathcal{N}=6$
states has already happened (see blow for the crossing rotational frequency).
For SLy4$_{\rm L}$ EDF, this crossing happens at a larger rotational frequency
($\hbar\omega\approx0.7$\,MeV) for this configuration.


\begin{figure*}
\centering
\includegraphics[scale=0.5]{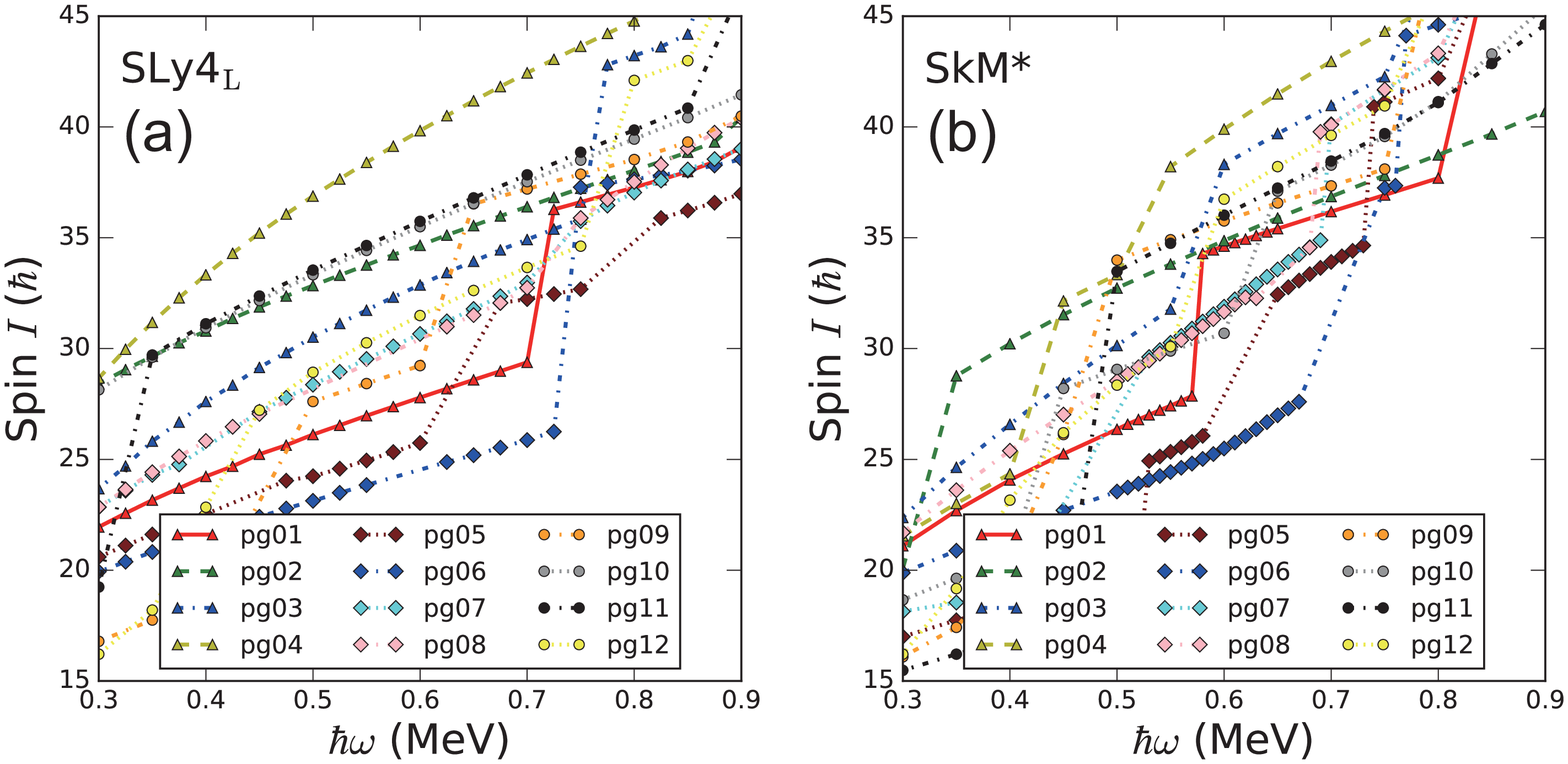}
\caption{The total angular momenta as a function of rotational frequency for different 
configurations at positive $\gamma$ values as a function of rotational frequency 
calculated with SLy4$_{\rm L}$ and SkM* EDFs. The solid, dashed, dotted, and dotted-dashed lines
denote states with ($\pi$,$\rho$)=(+,0), (+,+1), ($-$,0), and ($-$,+1), respectively.}
\label{figure3}
\end{figure*}

Figure~\ref{figure3} shows the total spin values for the configurations listed
in table~\ref{table1}, that are at the positive-$\gamma$ deformation (see
figure~\ref{figure2}), as a function of rotational frequency. It can be seen
that the angular momenta are in the range of 20-45\,$\hbar$ for
$\hbar\omega>0.5$\,MeV. For several configurations, a characteristic sudden
increase of angular momenta can be seen. This is due to the repel between the
$\mathcal{N}=6$ level, and the level below it in the same $(\pi,\rho)$-block at
$\hbar\omega\approx0.7$\,MeV, as shown in figure~\ref{figure1}. For SkM* EDF,
the alignment for the configuration `01' occurs at a smaller rotational
frequency, $\hbar\omega\approx0.6$\,MeV.

\begin{figure*}
\centering
\includegraphics[scale=0.5]{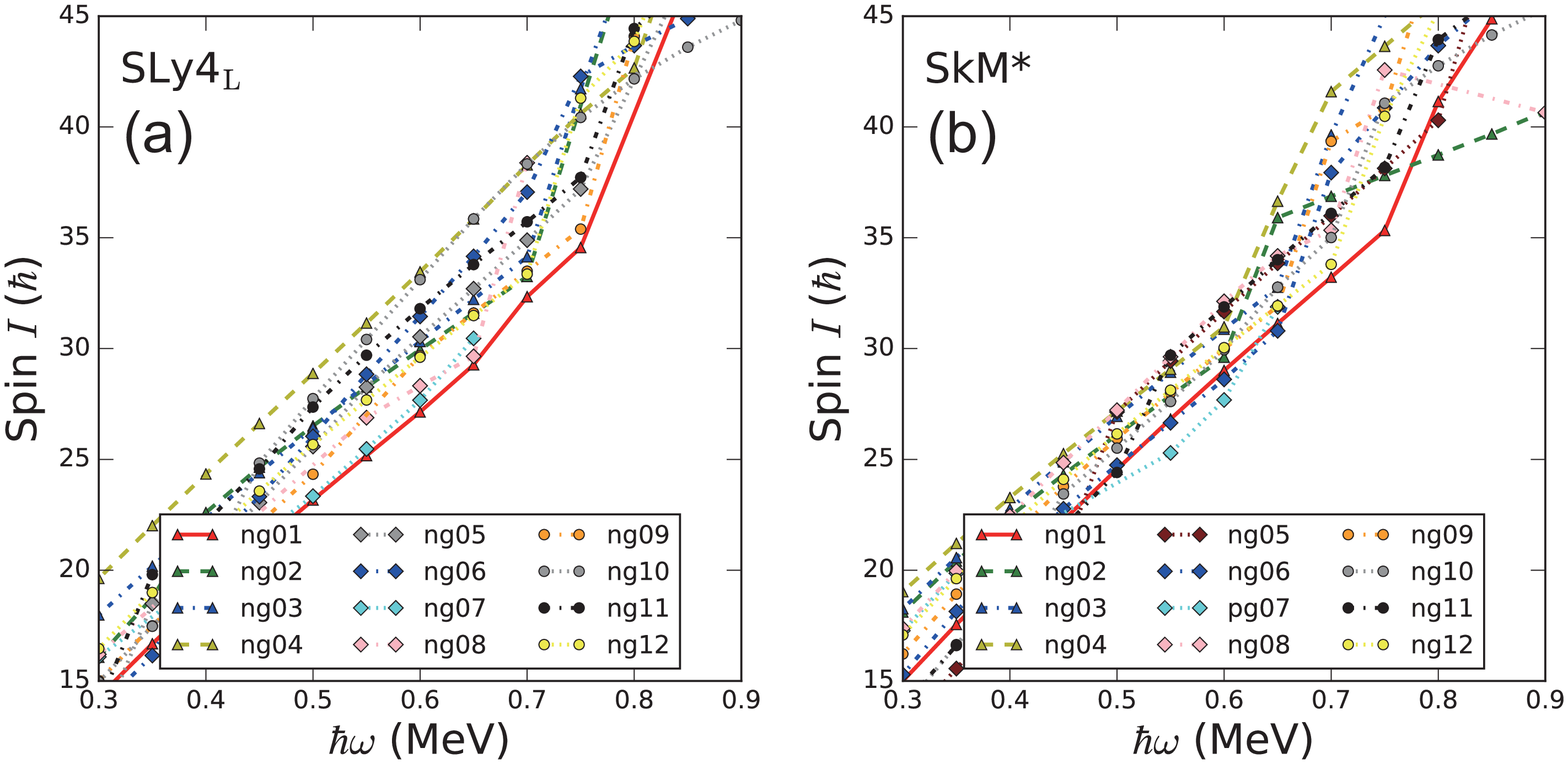}
\caption{The total angular momenta for different configurations at negative $\gamma$ 
values as a function of rotational
frequency calculated with SLy4$_{\rm L}$ and SkM* EDFs.}
\label{figure4}
\end{figure*}

In figure~\ref{figure4}, the same quantities as those in figure~\ref{figure3}
are plotted except that these states have negative $\gamma$ values. Compared to
the same configurations at positive-$\gamma$ deformation, the spin values for
all configurations, and for both EDFs, increase faster with increasing
$\hbar\omega$. In addition, the sudden increase of spin at certain $\omega$
values is less abrupt (or absent), as shown in figure~\ref{figure3}.

It has to be noted that, in this mean-field study, the total spin values are
determined by summing the individual contributions from occupied
single-particle levels. Hence, the total spin values are rather sensitive to
the calculated deformation and the specific configurations. This can be seen by
comparing figure \ref{figure3} and figure \ref{figure4}, which correspond
angular momenta variations as functions of rotational frequency at two
different deformations. It can be seen that the total spin values for the same
EDF, at the same frequency values, and for the same configuration can differ up
to $\sim$10$\hbar$. The variation as functions of rotational frequency is also
different for the same EDF, and the same configuration.

\begin{figure*}
\centering
\includegraphics[scale=0.5]{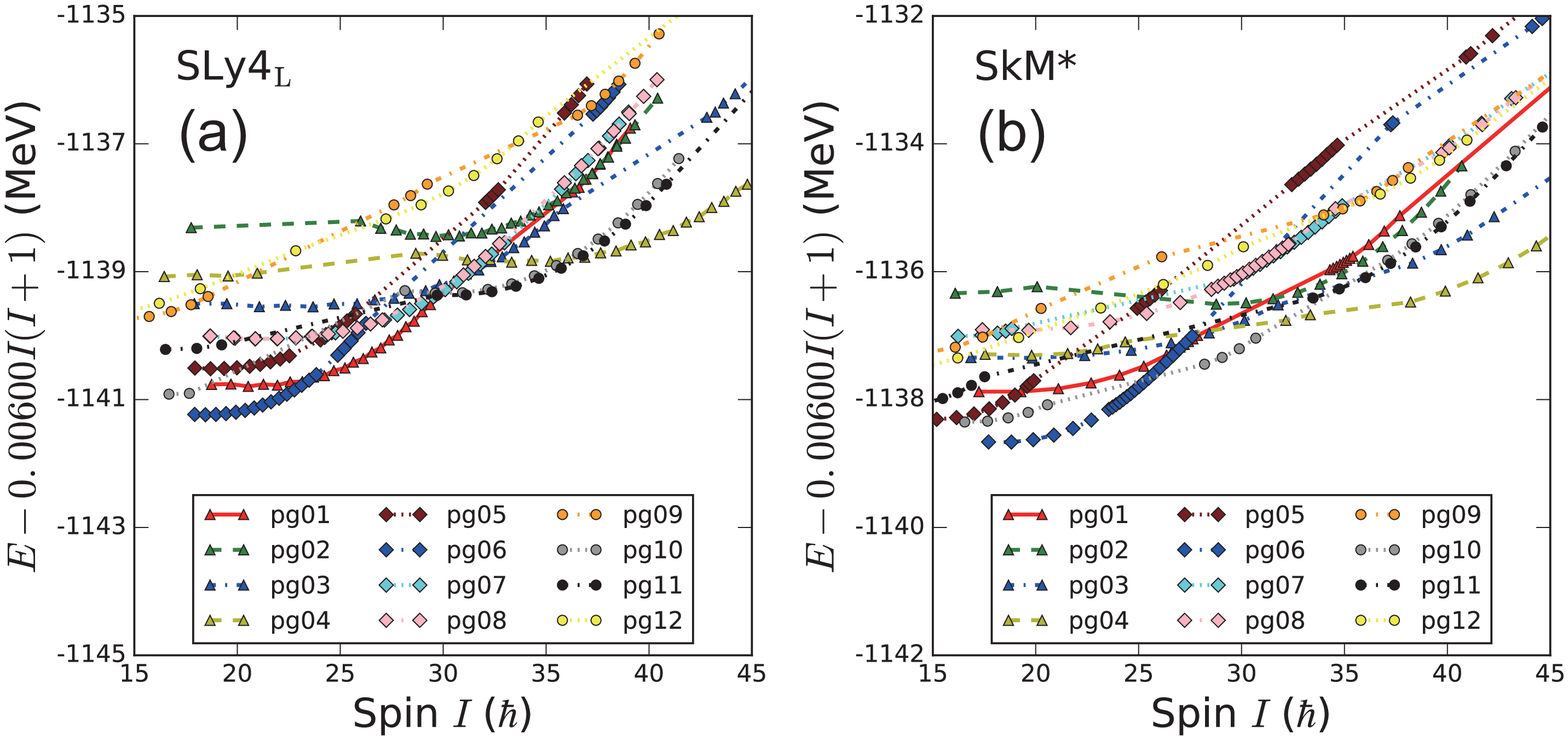}
\caption{Total energies for different configurations at positive $\gamma$ values. 
	Results are calculated with SLy4$_{\rm L}$ and SkM* EDFs.}
\label{figure5}
\end{figure*}

Figure~\ref{figure5} displays the total energies of $^{138}$Nd for
configurations listed in table~\ref{table1} at the positive-$\gamma$
deformation, calculated with SLy4$_{\rm L}$, and SkM* EDFs. For each
configuration, the energy of an arbitrary rotor has been subtracted from the
total energy. This is to bring the curves less steep for viewing the detail of
the curves more clearly. It can be seen that both EDFs predict similar trends
for same configurations. In addition, the alignments that have been seen in the
spin-versus-frequency figure (figure~\ref{figure3}) are reflected in these
energy-versus-spin curves, as a subtle change of the slopes. This can be
understood as the sudden increase of the moments of inertia which results in
the decrease of energies that are needed to provide the same amount of spin
increase.

\begin{figure*}
\centering
\includegraphics[scale=0.5]{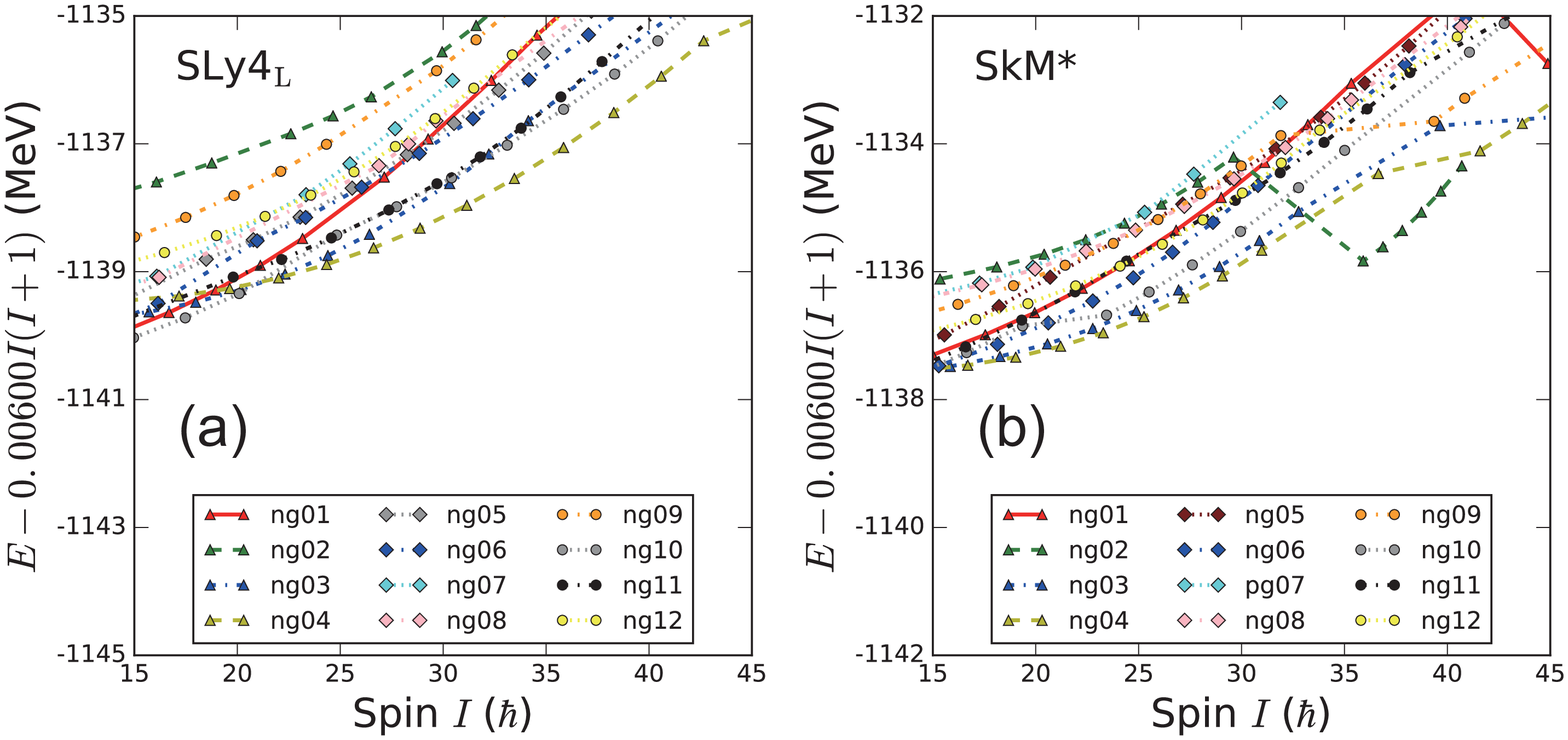}
	\caption{Same as figure~\ref{figure5}, except that the curves correspond to
	the configurations with negative $\gamma$ values.}
\label{figure6}
\end{figure*}

Figure~\ref{figure6} shows the same quantities as those in
figure~\ref{figure5}, except for negative-$\gamma$ values. It can be seen that
the changes due to the abrupt spin changes are absent in these bands. Comparing
the energies of the same configuration, it can be seen that the energies on the
negative-$\gamma$ side are in general 1-2\,MeV higher than that of their
positive-$\gamma$ states. For SkM* EDF, the `ng02' shows a large change.  This
is because the self-consistent calculation follows the positive-$\gamma$
minimum.

\begin{figure*}
\centering
\includegraphics[scale=0.45]{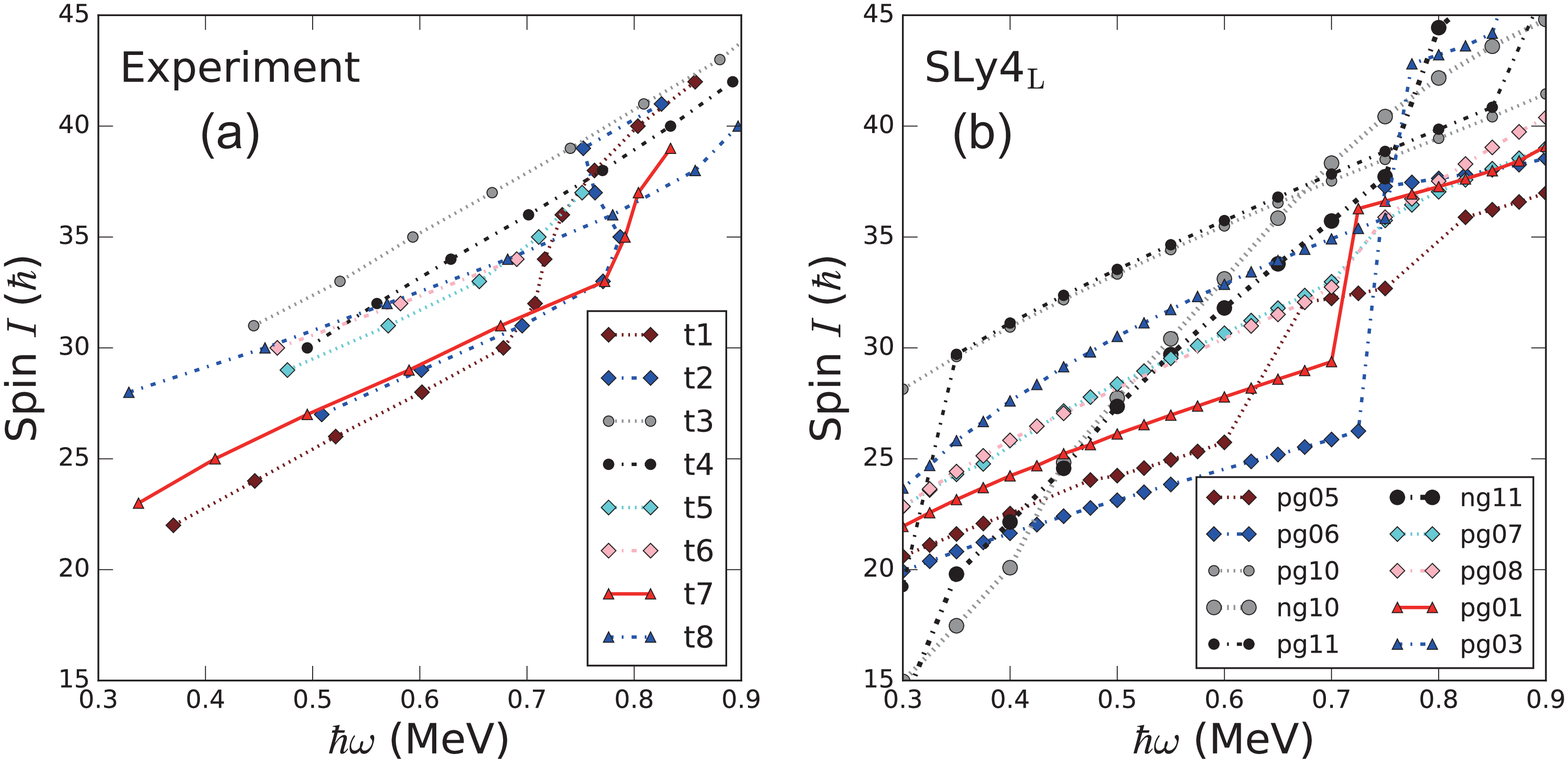}
\caption{Experimental angular momenta as a function of rotational frequency
(a) for `t1-t8' bands from Ref.~\cite{petr15}. The $\hbar\omega$ values for
$E(I)$ values are evaluated as $\frac{1}{2}E_{\gamma}(I+1\rightarrow
I-1)$. The calculated values (b) are extracted from
figures~\ref{figure3}(a) and \ref{figure4}(a). The proposed configuration assignment
of the current work for the experimental data are indicated by labelling 
the respective calculated curve with the same type of markers. 
See text for the case of `t3' and `t4' bands.}
\label{figure7}
\end{figure*}

Figure~\ref{figure7} compares the experimental spin values (a) of the
rotational bands, extracted from Ref.~\cite{petr15}, with the present cranked
DFT calculations using SLy4$_{\rm L}$ EDF (b). The calculated spin values are
extracted from figure~\ref{figure3}(a) and figure~\ref{figure4}(a). It can be
seen that the calculations display general consistencies with data in terms of
both the trend and the absolute values. In particular, both data and the
calculations show characteristic upbends at certain rotational frequency.
Specifically, the spin values of `T7' band upbend at
$\hbar\omega\approx0.75$\,MeV. The calculations underestimate this rotational
frequency by about 0.15 and 0.25\,MeV (see figure~\ref{figure3}) for SLy4$_{\rm
L}$(a) and SkM*(b) EDFs, respectively.

Note that this alignment in the present calculation is mainly due to the
crossing of the $\mathcal{N}=6$ state with the lower state in the same
$(\pi,\rho)$-block. It is rather sensitive to the parameter used, and in
particular, the predicted deformations of the minima. Hence, the critical
frequency at which the alignment occurs could be used as a criterion for
assessing the usefulness of a given parameter set used. One may use this
quantity to adjust parameterization if one needs to constrain on the
single-particle properties.

The transitions linking the high-spin states to states with lower spins have
been observed in Refs.~\cite{petr15}. This allows for the excitation energies
and the spin values to be determined. To compare these data with the
calculations, one needs to calculate the ground state of $^{138}$Nd where
pairing interaction is present. In table~\ref{table2}, I list the calculated
ground state total energies, and the pairing properties for $^{138}$Nd with
SLy4$_{\rm L}$ and SkM* EDFs. The calculation setup is identical to those used
to calculate the rotational bands, except that for the calculations of the
ground states, the pairing correlations are included through the
Skyrme-Hartree-Fock-Bogoliubov method~\cite{ring80}. The pairing cutoff
energies are $E_{\rm cutoff}=60$\,MeV, for both protons and neutrons. The pairing
strengths are determined to reproduce the experimental odd-even staggering
energies, which are 1.07, and 1.15\,MeV for neutrons, and protons,
respectively.

\begin{figure*}
\centering
\includegraphics[scale=0.5]{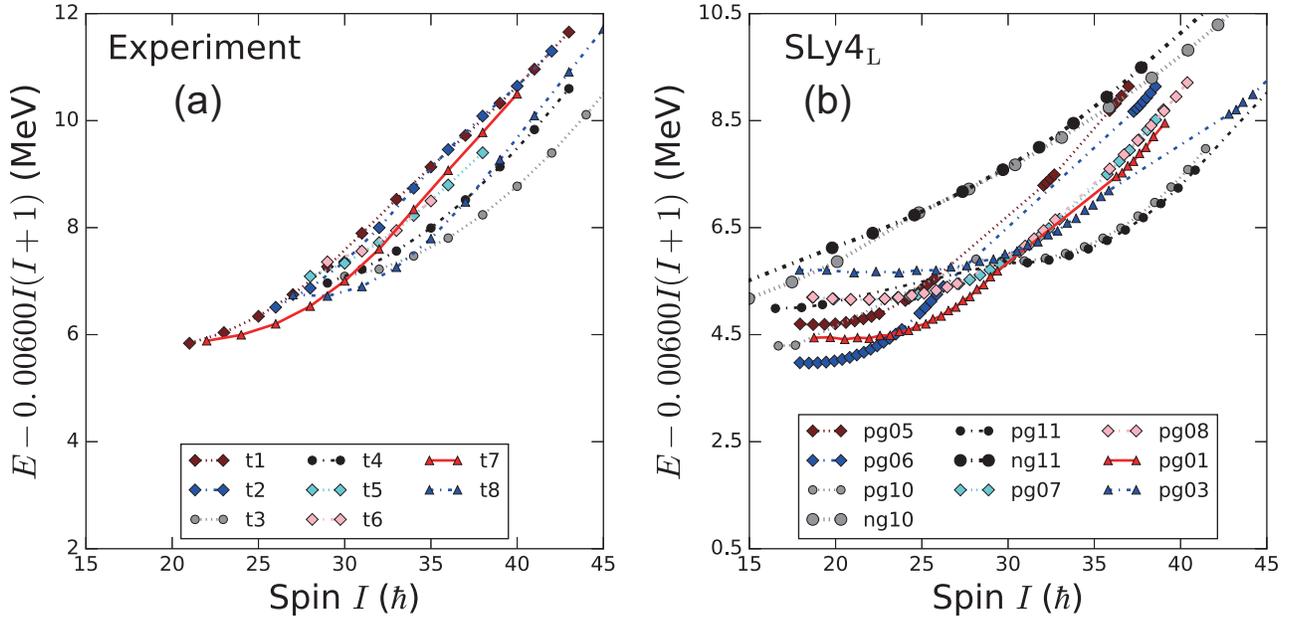}
\caption{Total energies as a function of angular momenta extracted from
experimental data~\cite{ange94,petr99,petr15}, compared with the SLy4$_{\rm L}$ shown in
figure~\ref{figure5}. The energy values are plotted with respect to the energy of the
22$^+$ state of `T7'. The calculated values (b) are extracted from
figures~\ref{figure5}(a) and \ref{figure6}(a).}
\label{figure8}
\end{figure*}

In figure~\ref{figure8}, I compare the observed excited energies~\cite{petr15}
(a) with the calculated ones with SLy4$_{\rm L}$ EDF, the latter of which are
extracted from figure~\ref{figure5}(a) and figure~\ref{figure6}(b). It can be
seen from figure~\ref{figure8} that the present calculations
[figure~\ref{figure8}(b)] systematically underestimate the observed excited
energies of the bands [figure~\ref{figure8}(a)] by $\sim1.5$\,MeV for
SLy4$_{\rm L}$ EDF. It should be noted that the calculations with SLy4$_{\rm
L}$ EDF predict prolate ground state, whereas the results using SkM* EDF
suggest a well deformed triaxial deformation for the ground state. One needs to
note that the current description may be insufficient for this nucleus near
ground state which has soft potential-energy surface. Additional effects such
as the vibrational motions may invite further correlations on the current
mean-field results. 

\begin{table}[htb]
\caption{The calculated total energies, pairing gaps, Fermi energies, and quadrupole moments
for the ground state of $^{138}$Nd using SHFB method.}
\label{table2}
\begin{ruledtabular}
\begin{tabular}{ccc}
& SLy4$_{\rm L}$ & SkM* \\
\hline
	$E_{\rm tot}$ (MeV) &  $-$1145.207  & $-$1141.870   \\
	$\Delta_n$ (MeV)    &  1.03     &  1.06  \\
	$\Delta_p$ (MeV)    &  1.06     &  1.10  \\
	$\lambda_n$ (MeV)   & $-$9.806  &  $-$9.696  \\
	$\lambda_p$ (MeV)   & $-$4.520  &  $-$4.018  \\
	$Q_{20}$ ({\it e}\,b) & 2.62      &  2.90 \\
	$Q_{22}$ ({\it e}\,b) & 0.00      &  1.15 \\
\end{tabular}
\end{ruledtabular}
\end{table}

Compared to the absolute energies, better indicators of a successful
description of rotational bands are (1) the relative energies between the
bands, and (2) the trend of the total energies as a function of spin values.
Taking into account of these two criteria [see figure~\ref{figure8}],
together with the spin-versus-$\omega$ plots [see figure~\ref{figure3}] allow
one to assign the experimentally observed bands~\cite{petr15} `T1', `T2', `T3',
`T4', `T5', `T6', `T7', and `T8' to the calculated ones with configurations
`pg05', `pg06', `pg10/ng10', `pg11/ng11', `pg07', `pg08', 'pg01', and `pg03',
respectively. For positive configurations other than those included in
table~\ref{table1}, the calculations predict rather high particle-hole
configurations. This is due to the opposite parity states of both protons and
neutrons for the single-routhian states, above and below the Fermi surfaces, as
shown in figure~\ref{figure1}. To create a positive one-particle-one-hole
state, a neutron or proton needs to be excited across the opposite states above
the Fermi surface, which results in high excitation energies.

There are a few interesting points worth mentioning along with the assignments.
First, the present calculations predict a yrast band with `pg01' configuration
for the spin values in the interval of $\sim$24-30$\hbar$. For spin values
lower than 24$\hbar$, configuration `pg06' is lower in energy. For higher spin
values ($I\ge30\hbar$), the yrast status of `pg01' is taken over by signature
partners `pg10,11'. This is consistent with the data, as shown in
figure~\ref{figure8}. For the near yrast states, the good agreement between the
calculations and the data~\cite{petr15} for spin values larger than 20$\hbar$
is important for evaluating the predictions of the current calculations. For
spin values between 20$\hbar$ and 25$\hbar$, the experimental information on
band `t2' is still missing, which prevents a direct comparison between
experimental data (`t2') and the calculations (`pg06'), see
figure~\ref{figure8}.

Second, a close examination of the experimental curve `T7' and the curve with
`pg01' configuration, reveals a change of the slope in the spin range of
30-35$\hbar$. This is, in the current calculation, interpreted to be due to the
alignment of neutrons at $\hbar\omega\approx0.7$\,MeV (see
figure~\ref{figure3}). The interpretation is similar for `T1', and `T2' bands,
where the calculations predict that the variations in the energy curves are due
to alignment. This differs from the interpretation in Ref.~\cite{petr13}, where
the macroscopic-microscopic calculations suggested a transition of the minimum
from a positive- to a negative-$\gamma$ deformation.

Third, if one assigns `T3', and `T4' to `pg10', and `pg11', respectively, then
one finds that the energy curves show slower increase with increasing spin
values compared to the experiment, as shown in figure~\ref{figure8}. The
calculated angular momenta seem to have a smaller slope compared to data, see
figure~\ref{figure7}. However, the energy curves and the spin curves
corresponding to the negative-$\gamma$ deformation seem to overestimate the
slopes of the curves of experimental data.

\begin{table}[htb]
\caption{Calculated proton qudrupole moments in $^{138}$Nd for different configurations,
with SLy4$_{\rm L}$ EDF. The `pg', and `ng' denote positive-, and 
negative-$\gamma$ minima, respectively.}
\label{table3}
\begin{ruledtabular}
\begin{tabular}{cccccccccc}
Band & $\hbar\omega$  & \multicolumn{2}{c}{$Q_{20}^p$ ({\it e}b)}  & \multicolumn{2}{c}{$-Q_{22}^p$ ({\it e}b)} & \multicolumn{2}{c}{$Q_t$ ({\it e}b)} & \multicolumn{2}{c}{$J$ ($\hbar$)} \\
\cline{3-4} \cline{5-6} \cline{7-8} \cline{9-10}
	&         (MeV)            &          pg   &               ng           &      pg      &           ng                &      pg        &         ng          &  pg & ng      \\
\hline
01  &   0.4   &  4.3  &  4.8  & 2.2 & $-$1.6 & 3.0 & 5.7 & 24.2 & 19.0 \\
    &   0.5   &  4.2  &  4.7  & 2.3 & $-$1.5 & 2.9 & 5.6 & 26.1 & 23.2 \\
    &   0.6   &  4.0  &  4.6  & 2.3 & $-$1.4 & 2.7 & 5.4 & 27.8 & 27.1 \\
    &   0.7   &  3.9  &  4.3  & 2.4 & $-$1.2 & 2.5 & 5.0 & 29.4 & 32.3 \\
    &   0.8   &  4.0  &  4.5  & 2.9 & $-$1.2 & 2.3 & 5.2 & 37.3 & 44.3 \\
\hline
05  &   0.4   &  3.8  &  4.6  & 2.0 & $-$1.3 & 2.6 & 5.4 & 22.5 & 20.8 \\
    &   0.5   &  3.6  &  4.5  & 2.1 & $-$1.1 & 2.4 & 5.1 & 24.2 & 25.6 \\
    &   0.6   &  3.5  &  4.2  & 2.2 & $-$0.9 & 2.2 & 4.7 & 25.7 & 30.5 \\
    &   0.7   &  3.4  &  4.0  & 2.3 & $-$0.8 & 2.1 & 4.6 & 32.2 & 34.9 \\
\hline
06  &   0.4   &  3.7  &  4.6  & 2.0 & $-$1.3 & 2.5 & 5.4 & 21.6 & 21.0 \\
    &   0.5   &  3.6  &  4.4  & 2.1 & $-$1.1 & 2.4 & 5.0 & 23.1 & 26.0 \\
    &   0.6   &  3.5  &  4.1  & 2.1 & $-$1.0 & 2.3 & 4.7 & 25.1 & 31.5 \\
    &   0.7   &  3.3  &  3.7  & 2.1 & $-$1.0 & 2.1 & 4.3 & 25.9 & 37.1 \\
    &   0.8   &  3.3  &  2.5  & 1.8 & $-$1.4 & 2.3 & 3.3 & 37.6 & 43.7 \\
\hline
10  &   0.4   &  4.7  &  4.1  & 2.2 & $-$1.4 & 3.4 & 4.9 & 30.9 & 20.1 \\
    &   0.5   &  4.5  &  4.3  & 2.3 & $-$1.1 & 3.2 & 4.9 & 33.3 & 27.7 \\
    &   0.6   &  4.4  &  4.0  & 2.4 & $-$1.0 & 3.0 & 4.6 & 35.5 & 33.1 \\
    &   0.7   &  4.2  &  3.6  & 2.4 & $-$1.1 & 2.8 & 4.2 & 37.5 & 38.3 \\
    &   0.8   &  4.1  &  3.2  & 2.5 & $-$1.1 & 2.7 & 3.8 & 39.5 & 42.2 \\
\hline
11  &   0.4   &  4.7  &  4.6  & 2.2 & $-$1.3 & 3.4 & 5.4 & 31.1 & 22.2 \\
    &   0.5   &  4.5  &  4.4  & 2.3 & $-$1.1 & 3.2 & 5.0 & 33.5 & 27.4 \\
    &   0.6   &  4.4  &  4.1  & 2.4 & $-$1.0 & 3.0 & 4.7 & 35.7 & 31.8 \\
    &   0.7   &  4.2  &  3.9  & 2.4 & $-$0.9 & 2.8 & 4.4 & 37.8 & 35.7 \\
    &   0.8   &  4.0  &  3.6  & 2.5 & $-$1.0 & 2.6 & 4.2 & 39.9 & 44.4 \\
\end{tabular}
\end{ruledtabular}
\end{table}

To verify the second and third interpretations raised by the current
calculations, one needs to conduct experimental measurements of the
transitional quadrupole moments or life time on the bands `T1,2,3,4,7', as well
as a few other states for references. Table~\ref{table3} lists the transitional
quadrupole moments calculated with SLy4$_{\rm L}$ at both positive- and
negative-$\gamma$ values for a few configurations. Incidentally, a recent
experiment~\cite{bell18} performed interesting life-time measurement on the
rotational bands near ground state of $^{138}$Nd. The current work suggests
that the life-time data in the spin intervals of 30-40$\hbar$ are highly
desirable. A comparative study among the bands with spin increase, together
with theoretical calculations will reveal better modelings, better
parameterizations, as well as suitable testing grounds for phenomena such as
shape transitions, rotational axis re-orientations, alignments with triaxial
shape, all in the same nucleus.

\section{summary} 
\label{summary}

In the current work, cranked self-consistent Skyrme-Hartree-Fock (SHF)
calculations were applied to the description of the rotational bands recently
observed in $^{138}$Nd in the spin range of 20$-$45$\hbar$. For the
configurations near yrast line, the calculations with both SkM* and SLy4$_{\rm
L}$ parameters predicted pronounced triaxial minima at positive-$\gamma$
deformation and comparatively soft triaxial minima at negative-$\gamma$
deformation. The configurations with positive-$\gamma$ deformation were
calculated to be lower in energy compared to those with negative-$\gamma$
deformation for both parameters.

By comparing the calculated total angular momenta as a function of rotational
frequency and the relative total energies as a function of spin values for
different configurations with experimental data, the configurations of the
observed bands~\cite{petr15} `T1$-$8' were assigned in the current cranked SHF
calculations. It needs to be noted that for some configurations, the current
interpretations of structural changes differ from those of previous
macroscopic-microscopic calculations~\cite{petr15}. The predicted transitional
quadrupole moments were provided for future life-time measurement to verify the
current interpretations.

\begin{acknowledgments}

The current work is supported by National Natural Science Foundation of
China (Grant No. 11705038).
I thank the HPC Studio at Physics Department of Harbin Institute of 
Technology for computing resources allocated through INSPUR-HPC@PHY.HIT.

\end{acknowledgments}

\bibliographystyle{apsrev4-1}
%

\end{document}